\documentclass[10pt,superscriptaddress,twocolumn,amsmath,amssymb,aps,prl,showpacs]{revtex4-1}
\usepackage{mathrsfs}
\usepackage{graphicx}
\usepackage{dcolumn}
\usepackage{bm}

\usepackage{graphicx}
\usepackage{epstopdf}

\usepackage{amsmath,amssymb,mathrsfs}
\usepackage{braket}
\usepackage{hyperref}
\hypersetup{
    colorlinks=true,
    linkcolor=blue,
    filecolor=blue,      
    urlcolor=blue,
    citecolor=blue
    }

\begin{document}


\title{Quantum Spin Dynamics in a Normal Bose Gas with Spin-orbit Coupling}


\author{Wai Ho Tang}
\affiliation{Department of Physics and Center of Theoretical and Computational Physics, The University of Hong Kong, Hong Kong, China}

\author{Shizhong Zhang}
\email{shizhong@hku.hk}
\affiliation{Department of Physics and Center of Theoretical and Computational Physics, The University of Hong Kong, Hong Kong, China}

\date{\today}

\begin{abstract}
In this Letter, we investigate spin dynamics of a two-component Bose gas with spin-orbit coupling realised in cold atom experiments. We derive coupled hydrodynamic equations for number and spin densities as well as their associated currents. Specialising to quasi-one-dimensional situation, we obtain analytic solutions of the spin helix structure and its dynamics in both adiabatic and diabatic regimes. In the adiabatic regime, the transverse spin decays parabolically in the short-time limit and exponentially in the long time limit, depending on initial polarisation. In contrast, in the diabatic regime, transverse spin density and current oscillate in a way similar to the charge-current oscillation in an undamped LC circuit. The effects of Rabi coupling on the short-time spin dynamics is also discussed. Finally, using realistic experimental parameters for $^{87}$Rb, we show that the time scales for spin dynamics is of order of milliseconds to a few seconds and can be observed experimentally.
\end{abstract}

\maketitle


{\em Introduction.} It has long been recognised that collective spin dynamics of quantum mechanical origin can exist in a dilute gas at temperature $T\gtrsim T_d$, where $T_d$ is the degeneracy temperature. It arises due to indistinguishability of identical atoms in binary scattering and is known as the identical spin rotation effect (ISRE)~\cite{jphys.43.197, PhysRevLett.52.1508, PhysRevLett.52.1512}. This effect is operative for both bosons \cite{PhysRevLett.88.230403, PhysRevLett.88.230404, PhysRevLett.88.230405} and fermions~\cite{PhysRevA.79.051601, PhysRevLett.102.215301} and has led to the observations of spin waves and anomalous spin segregation for weakly interacting bosons~\cite{PhysRevLett.88.070403} and fermions~\cite{PhysRevLett.101.150401}. Similar effects also occurs in a degenerate Fermi liquid like $^3$He  where it leads to anomalous spin diffusion known as the Leggett-Rice effect~\cite{PhysRevLett.20.586, 0022-3719-3-2-027}. Recently, Leggett-Rice effect has also been observed in unitary Fermi gas in both two~\cite{nphys2637} and three dimensions~\cite{Bardon722, PhysRevLett.114.015301}.

The ISRE effects explored so far are limited to systems with spin-SU$(2)$ symmetry where the total spin is a good quantum number and its dynamics decouples from that of the density~\cite{PhysRevLett.88.230403, PhysRevLett.88.230404, PhysRevLett.88.230405,PhysRevA.79.051601, PhysRevLett.102.215301}.  In this Letter, we investigate the spin dynamics of a normal Bose gas with spin-orbit coupling (SOC) that was recently realized in cold atom experiments \cite{nature09887, PhysRevLett.109.095301, PhysRevLett.109.095302, PhysRevLett.109.115301, PhysRevLett.111.095301, PhysRevA.88.021604, PhysRevA.90.013616, ncomms5023, PhysRevX.6.031022, PhysRevA.94.061604, nphys3672, PhysRevLett.117.185301, PhysRevLett.117.235304, science.aaf6689}. The coupling between spin and orbit degrees of freedom breaks the SU$(2)$ symmetry and leads to more intricate dynamics that has no analog in usual dilute gases discussed above. In particular, we show how the long-wavelength and low-frequency hydrodynamic equations are modified in the presence of SOC, and how it leads to the appearance of persistent spin helical (PSH) structure. The decay of the spin helical structure is discussed in both adiabatic and diabatic limits. The general equations we obtain should serve as the starting point for investigating spin dynamics in a spin-orbit coupled Bose gas such as spin waves and their attenuations. Spin dynamics for Fermi gas with spin-orbital coupling has been discussed in Refs.\cite{PhysRevLett.99.110403, PhysRevA.87.041602, PhysRevB.87.125416, PhysRevA.88.033613, PhysRevA.88.043634, PhysRevLett.112.095302, PhysRevA.92.013607, PhysRevA.93.063635}.


{\em General Setup.} For definiteness, let us consider a gas of bosonic atoms $^{87}\mathrm{Rb}$ of mass $m$ with two hyperfine-Zeeman sub-levels $\ket{F,m_F} \equiv \ket{1,0} \equiv \ket{\uparrow}$ and $\ket{1,-1}\equiv \ket{\downarrow}$ that are coupled by a pair of Raman lasers with momentum transfer $\mathbf{q}=q\hat{\mathbf{x}}$ along the $\hat{\mathbf{x}}$-direction. We set the two-photon detuning to be zero for simplicity in the following discussion. The harmonic trapping potential $V(\mathbf{r})$, independent of spin, is assumed to be strong in the $\hat{\mathbf{y}}$- and $\hat{\mathbf{z}}$-directions but weak in the $\hat{\mathbf{x}}$-direction and the system can be considered quasi-one-dimensional. The $s$-wave interaction is almost SU$(2)$ invariant for $^{87}${Rb} and is given by a single coupling constant $g$. The Hamiltonian can be written as $\hat{\mathscr{H}}=\int d^3 \mathbf{r} \sum_{\mu,\nu=\uparrow,\downarrow}\psi^\dagger_{\mu}(\mathbf{r})H_{\mu\nu}\psi_{\nu}(\mathbf{r})+\frac{1}{2}g\int d^3\mathbf{r} \colon\hat{n}(\mathbf{r})\hat{n}(\mathbf{r})\colon$ where $H_{\mu\nu}$ is given by
\begin{equation}
H_{\mu\nu}=\left[\dfrac{-\hbar^2 \nabla^2}{2m} + V(\mathbf{r})\right]\delta_{\mu\nu} -\frac{i\hbar q}{m}\sigma^z_{\mu\nu} \partial_x + \frac{\hbar\Omega_R}{2}\sigma^x_{\mu\nu}~.
\end{equation}
$\hat{\psi}_{\mu}(\mathbf{r})$ ($\hat{\psi}^\dagger_{\mu}(\mathbf{r})$) is the annihilation (creation) operator for boson with spin $\mu$ at position $\mathbf{r}$. $\Omega_R$ is the two-photon Rabi coupling. The number and spin densities are then given by $\hat{n}(\mathbf{r})=\sum_{\mu}\hat{\psi}^\dagger_{\mu}(\mathbf{r})\hat{\psi}_{\mu}(\mathbf{r})$ and ${\hat{s}}_i(\mathbf{r}) = \frac{1}{2}\sum_{\mu,\nu}\hat{\psi}^\dagger_\mu(\mathbf{r})\sigma^i_{\mu\nu}\hat{\psi}_\nu(\mathbf{r})$, respectively. $\sigma^i$ are the Pauli matrices. In what follows, we use arrow on top of an operator to indicate that it is a vector in spin space while {boldface $\hat{\mathbf{x}}, \hat{\mathbf{y}}, \hat{\mathbf{z}}$} describes the spatial direction. Properties of condensation described by $\hat{\mathscr{H}}$ have been discussed extensively in the literature, including its phase diagram and collective excitations~\cite{PhysRevLett.105.160403, PhysRevLett.107.150403, doi:10.1142/S0217979212300010, PhysRevLett.108.225301, PhysRevLett.110.235302, nphys2905, PhysRevLett.114.105301, 0034-4885-78-2-026001, doi:10.1142/9789814667746_0005, srep15307, Zhang2016, PhysRevLett.118.145302, PhysRevLett.120.120401} as well as spin dynamics~\cite{PhysRevA.93.063420}.

{\em Transport Equations.} We first derive the continuity equations for number and spin densities and also identify the modifications to the associated number and spin currents due to spin-orbit coupling. We restrict ourselves to transport along the $\hat{\mathbf{x}}$-direction. Using Heisenberg's equation of motion $i\hbar \braket{\partial_t \hat{A}} = \braket{[\hat{A}, \hat{\mathscr{H}}]}$ with $\hat{A}$ being the number $\hat{n}(\mathbf{r})$ and spin $\hat{\vec{s}}(\mathbf{r})$ densities, we find immediately
\begin{equation}
\partial_t n + \partial_x j_0 = 0~\label{eq:DEOM}
\end{equation}
where the number current along $\hat{\mathbf{x}}$-direction $j_0=\langle\hat{j}_0\rangle$ with $\hat{j}_0=-{i\hbar}/({2m})\sum_{\mu}(\hat{\psi}^\dagger_{\mu}\partial_x \hat{\psi}_{\mu}-\partial_x\hat{\psi}^\dagger_{\mu}\hat{\psi}_{\mu})+ (2q/m)\hat{s}_z$. We note that due to SOC, the number current is coupled to the $\hat{z}$-component of the spin density. This redefinition is recently found to cause the violation of irrotationality of velocity field in spin-orbit couple condensate and the reduction of the quantum of circulation \cite{PhysRevLett.118.145302}. 

In the presence of spin-orbit coupling, the total spin is no longer conserved and the definition of spin current operator $\hat{\vec{j}}$ is not entirely obvious. In our case, we identify the spin current by grouping all the gradient term in the continuity equation for spin density
\begin{equation}
\partial_t \vec{s} + \partial_x \vec{j} = \Omega_R \hat{x}\times\vec{s} + (2q/\hbar)\hat{z}\times\vec{j}~\label{eq:SEOM}
\end{equation}
where the spin current operator is given by $\hat{\vec{j}}=-{i\hbar}/({4m})\sum_{\mu,\nu}\hat{\psi}^\dagger_\mu\vec{\sigma}_{\mu\nu}\,\partial_x \hat{\psi}_\nu + \mathrm{H.c.}+\hat{n}q/(2m)\hat{z}$. We note two important modifications due to SOC. Firstly, the spin-current is now coupled to the total density of the system. Secondly, apart from the usual spin precessing term due to Rabi coupling, there is an additional precessing term, proportional to the strength of SOC, of spin current along $\hat{z}$-direction in Eq.(\ref{eq:SEOM}). We note that the modified definition of spin current operator $\hat{\vec{j}}$ can also be motivated from semiclassical considerations. Let the distribution function (a matrix in spin space) be given by $\hat{f}(\mathbf{r},\mathbf{p},t)$, then one can define the spin current as 
\begin{equation}
\vec{j}(\mathbf{r},t)=\frac{1}{2}{\rm Tr}\int \frac{d^3 \mathbf{p}}{(2\pi \hbar)^3}\hat{f}(\mathbf{r},\mathbf{p},t)\frac{1}{2}\left(\vec{\sigma}\frac{\partial H}{\partial p_x}+\frac{\partial H}{\partial p_x}\vec{\sigma}\right)~.
\end{equation}
The symmetrization is necessary because of non-commutivity of $\vec{\sigma}$ and ${\partial H}/{\partial p_x}$. Since ${\partial H}/{\partial p_x}=p_x/m+(q/m)\hat{\sigma}^z$. The first term $p_x/m$ corresponds to the standard spin-current operator, while the second term $(q/m)\hat{\sigma}^z$ only modifies the $\hat{z}$-component of the spin-current  by an additional term $\hat{n}q/(2m)$.

Using the operator forms of the number and spin currents, it is now straightforward to obtain their equations of motions, which are much more complicated because of the involvement of the momentum flux tensors. However, in the normal state above the degeneracy temperature, the momentum flux tensors can be simplified using Boltzmann distribution (recall $T\gtrsim T_d$) and gradient expansion (for detailed derivation, see Supplemental Material~\cite{supp}). As a result, we obtain
\begin{align}
\partial_t j_0+\dfrac{k_BT}{m} \partial_x n &=\frac{2q}{m}\Omega_R s_y-\frac{g}{2m}\partial_x\left(\frac{3}{4}n^2+\vec{s}^{\,2}\right)\label{eq: current density EOM}\\ 
\partial_t \vec{j} + \alpha \partial_x \vec{s} &= \left(\Omega_R \hat{x} + \frac{g}{\hbar} \vec{s}\right)\times\vec{j} +\frac{2q\alpha}{\hbar}\hat{z}\times\vec{s}\nonumber \\& +\frac{qn\Omega_R}{2m}\hat{y} - \frac{3g}{4m} (\partial_x n)\vec{s} - \gamma \vec{j}~,\label{eq: spin current density EOM}
\end{align}
where $\alpha = {k_B T}/{m} + {ng}/(4m) $. A phenomenological spin current relaxation term $-\gamma\vec{j}$ is added to Eq.(\ref{eq: spin current density EOM}). In the absence of the spin-orbit coupling ($\Omega_R=0$ and $q=0$), Eqs.(\ref{eq:SEOM},\ref{eq: spin current density EOM}) reduce to the standard Leggett-Rice form for a degenerate Fermi liquid \cite{PhysRevLett.20.586, 0022-3719-3-2-027}. It is noteworthy that the spin gradient term ${ng}/({4m})\partial_x \vec{s}$ in Eq.(\ref{eq: spin current density EOM}) is usually omitted in comparison to the Leggett-Rice term $(g/\hbar)\vec{s}\times\vec{j}$ when the spatial variation of $\vec{s}$ is small. In the presence of SOC, however, it has to be retained because the natural scale of variation for $\vec{s}$ is set by the spin-orbit scale $q$ which can be quite large. In addition, due to the fast temporal variation of spin density, it is necessary to go beyond the adiabatic approximation $|\partial_t \vec{s}/\vec{s}| \lesssim \gamma$ usually assumed in literature and discuss the dynamics in the diabatic regime as well. 

Equations (\ref{eq:DEOM},\ref{eq:SEOM},\ref{eq: current density EOM},\ref{eq: spin current density EOM}) form the basic equations for the spin dynamics of a SOC boson above the degeneracy temperature. In following, we first discuss the limit when the effects of Rabi coupling $\Omega_R$ is small, or what is equivalent, for time $t\ll 1/\Omega_R$, and discuss the existence of persistent spin helix (PSH) at wave vector $k=2q$ (hereafter $\hbar = 1$) and its decay when $k$ deviates from $2q$. The effects of Rabi term on PSH will be discussed in the end of the Letter.

{\em {Persistent spin helical structure.}} The full set of equations allow an exact solution corresponding to persistent spin helix with uniform density $n=n_0$, spin density $s_z=s_{z,0}$ and $\vec{s}^{\,2}\equiv\vec{s}\cdot\vec{s}$ that are independent of time. If we write the transverse spin $\vec{s}_\perp = s_x\hat{x} + s_y\hat{y}$ in terms of $s^\pm = s_x \pm i s_y$, and likewise for the spin currents $\vec{j}(x,t) = \vec{j}_\perp(x,t) + j_{z}(t)\hat{z}$. Then for the spin helical structure with definite wave number $k$, we can write $s^{\pm}(x,t) = e^{\pm ikx}\tilde{s}^{\pm}(t)$ and similarly $j^{\pm}(x,t)=e^{\pm ikx}\tilde{j}^{\pm}(t)$ and obtain the following set of equations
\begin{align}
\partial_t \tilde{s}^+ &= - i (k-2q)\tilde{j}^+~,\label{ds tilde}\\
\partial_t \tilde{j}^+  &= (i \lambda s_{z,0} - \gamma) \tilde{j}^+ -i[\alpha(k-2q) + \lambda j_z]\tilde{s}^+,\label{dj tilde}\\
\partial_t j_z &= \lambda {\rm Im}[\tilde{s}^- \tilde{j}^+] -\gamma j_z,\label{djz}
\end{align}
where $\lambda = g/\hbar$ and ${\rm Im}$ denotes the imaginary part. When $k=2q$, the transverse spin $\tilde{s}^+$ is independent of time and corresponds to a static spin helical structure in which spin density rotates about $\hat{z}$-axis with wave vector $2q$ in the $\hat{\mathbf{x}}$ direction,
\begin{equation}
\vec{s}_{\rm psh} = s_{\perp,0} \cos(2qx)\hat{x} + s_{\perp,0} \sin(2qx)\hat{y} + s_{z,0}\hat{z}~.
\end{equation}
In semiconductor heterostructure, it is understood that the persistent spin helix is due to an emergent SU$(2)$ symmetry in the presence of spin-orbit coupling~\cite{PhysRevLett.97.236601, nature07871, RevModPhys.89.011001}. In the long time limit $t\gg 1/\gamma$, it is easy to see that both $j_z$ and $\tilde{j}^+$ decays to zero, according to Eqs.(\ref{dj tilde},\ref{djz}).

\begin{figure}[h]
\includegraphics[width=85mm,scale=1]{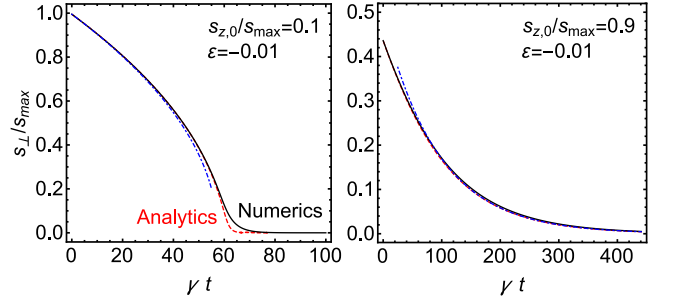}
\caption{\label{fig:AdiaPt}(Colour online) Short- and long-time behaviours of the transverse spin in the adiabatic regime. Short-time decay for initial spin density polarised close to the $xy$-plane. The decay is parabolic (left panel). Long-time decay for initial spin density polarised close to the $z$-axis. The decay is exponential (right panel). Numerical simulations of the full set of Eqs. (\ref{eq:DEOM},\ref{eq:SEOM},\ref{eq: current density EOM},\ref{eq: spin current density EOM}) (black solid) agree very well with the analytical results (red dashed) given by Eq.(\ref{eq: sperp}). Blue dashed lines shows the asymptotic results, Eqs.(\ref{asym eq para}, \ref{asym eq exp}). The deviation at the tail of left panel between the simulation and analytic equation (\ref{eq: sperp}) indicates the failure of adiabatic approximation. }
\end{figure}

{\em {Vicinity of PSH.}} In the following, we investigate the dynamics of spin helical structure when its wave vector deviates away from $2q$, described by the parameter $\varepsilon\equiv k/(2q)-1$. Here it is important to distinguish two regimes. In the adiabatic regime where the spin currents can relax much faster than the spin densities and can thus follow adiabatically the time evolution of spin density, $|\partial_t \vec{s}/\vec{s}\,| \lesssim \gamma$, we can set $\partial_t \tilde{j}^+ = \partial_t j_z = 0$ in Eqs.(\ref{dj tilde},\ref{djz}) in the steady state. Writing $\tilde{s}^+(t)\equiv s_\perp(t)\exp[i\theta(t)]$, we obtain the following set of equations
\begin{align}
\begin{split}
(\gamma^2 + \lambda^2 s_{z,0}^2)\ln\dfrac{s_\perp(t)}{s_{\perp,0}} + \frac{\lambda^2}{2}\left[s_{\perp}^2(t) - s_{\perp,0}^2\right] \\= -\alpha\gamma (k - 2q)^2 t~\label{eq: sperp},
\end{split}\\
\theta(t) &= \dfrac{\lambda s_{z,0}}{\gamma}\ln\left[\dfrac{s_\perp(t)}{s_{\perp,0}}\right]~,\\
j_z &= -\dfrac{\lambda s_\perp^2 \alpha(k-2q)}{\gamma^2 + \lambda^2 (s_\perp^2 + s_{z,0}^2)}~,\\
\tilde{j}^+ &= \tilde{s}^+ \dfrac{\alpha(k-2q)(\lambda s_{z,0} - i \gamma)}{\gamma^2 + \lambda^2 (s_\perp^2 + s_{z,0}^2)}~,\label{adia jtilde}
\end{align}
where $s_{\perp,0}=s_\perp(t=0)$. Substitution of $k=2q$ recovers the previous solution of PSH. When $k \neq 2q$, the transverse spin magnitude decays according to Eq.(\ref{eq: sperp}). Depending on the relative magnitude of $s_\perp$ and $s_{z,0}$, one can distinguish two qualitatively different behaviours. \\
(i) When $|s_{\perp,0}| \geq |s_\perp(t)| \gg |s_{z,0}|$, namely, when spins are polarized close to the $xy$-plane, the first term on the left of Eq.(\ref{eq: sperp}) is negligible, hence the transverse spin magnitude decays parabolically in the short time limit $t\ll \tau_{\rm para}$,
\begin{equation}
s_\perp (t) \approx s_{\perp,0}\sqrt{1 - \frac{t}{\tau_{\rm para}}}, ~~\tau_{\rm para} = \frac{\lambda^2 s_{\perp,0}^2}{2\alpha\gamma(k-2q)^2 }, \label{asym eq para}
\end{equation}
where the time constant $\tau_{\rm para}$ depends quadratically on the interaction parameter $\lambda$ and inversely on the spin current relaxation rate $\gamma$. As expected, it diverges when $k=2q$. \\
(ii) In the long time limit $t\gg \tau_{\rm para}$ when $|s_\perp(t)| \ll |s_{z,0}|$, the decay becomes exponential in the adiabatic regime with a different time constant $\tau_{\rm exp}$
\begin{equation}
s_\perp (t) \approx s_{\perp,0} e^{-t/\tau_{\rm exp}},~~~\tau_{\rm exp} = \frac{\gamma^2 + \lambda^2 s_{z,0}^2}{\alpha\gamma(k-2q)^2} \label{asym eq exp}.
\end{equation}
Fig.\ref{fig:AdiaPt} shows the excellent agreement between above analytic formulae and numerical results. We note that the dynamical equation (\ref{eq: sperp}) is similar in form to the Leggett-Rice equation derived for a degenerate Fermi liquid~\cite{0022-3719-3-2-027}, except for the explicit appearance of spin-orbit coupling term on the right hand side of Eq.(\ref{eq: sperp}). We emphasize that Eqs.(\ref{asym eq para}, \ref{asym eq exp}) apply so long as $1/\Omega_R\gg \tau_{\rm para}, \tau_{\rm exp}$ even in the presence of a small Rabi coupling.


{\em {Diabatic Regime.}} When the wave vector $k$ of the spin helix deviates significantly away from the PSH wave vector $2q$, the adiabatic condition fails. In the diabatic regime when $|\partial_t \vec{s}/\vec{s}| \gg \gamma$, we can neglect $\lambda |j_z|$ in Eq.(\ref{dj tilde}), and as a result $\tilde{s}^+(t), \tilde{j}^+(t)$ form a closed dynamical system in Eqs.(\ref{ds tilde},\ref{dj tilde}). With the initial condition $(\tilde{s}^+, \tilde{j}^+, j_z)_{t=0} = (s_{\perp,0},0,j_{z,0})$, one obtains~\cite{supp} 
\begin{align}
\tilde{s}^+(t)=& s_{\perp,0} e^{-\frac{\gamma t}{2} + i \frac{\lambda s_{z,0}}{2} t } \left[\cos \Gamma t + \dfrac{\gamma - i \lambda s_{z,0}}{2\Gamma}\sin \Gamma t \right] ,\label{general stilde}\\
\tilde{j}^+(t)=& i\sqrt{\alpha} s_{\perp,0} e^{-\frac{\gamma t}{2} + i \frac{\lambda s_{z,0}}{2} t}\sin(\Gamma t)\,\mbox{sign}(2q - k),\label{general jtilde}\\
j_z(t)=& j_{z,0} e^{-\gamma t} \nonumber - e^{-\gamma t} \dfrac{\lambda s_{\perp,0}^2}{4 (k-2q)}\\&\times \left[ 1 + \gamma t - \cos(2\Gamma t) - \dfrac{\gamma}{2}\dfrac{\sin(2\Gamma t)}{\Gamma} \right]~,
\end{align}
where $\Gamma=\sqrt{\alpha}|k-2q|$. In obtaining the above simplified expressions we have assumed that the polarization of spin are close to the $xy$-plane and as a result $\Gamma \gg \gamma, \lambda s_{z,0}$.

\begin{figure}[h]
\includegraphics[width=85mm,scale=1]{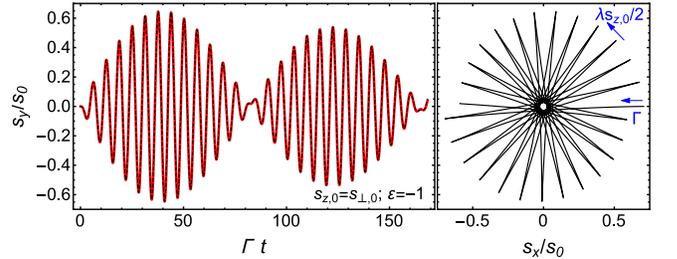}
\caption{\label{fig:DiaSX}(Colour online) Left panel shows time dependence of $y$-component of spin density for spin density polarized along $\hat{x}$-direction at $t=0$. Numerical result (black dashed) agrees very well with the analytical result (red solid), Eq.(\ref{general stilde}). Right panel is the trajectory of transverse spin component in the $xy$-plane. Also indicated in the graph are the fast oscillations of the magnitude of transverse spin ($\Gamma$) and its slow rotations with rate $\lambda s_{z,0}/2$.}
\end{figure}

The dynamics of the transverse components consists of three parts: fast oscillation in magnitude with frequency $\Gamma$, slow precessing of the axis of oscillation with frequency $\lambda s_{z,0}/2$ and the damping of oscillation amplitude with the rate of $\gamma/2$, as shown in Fig.\ref{fig:DiaSX}. If one neglects the small correction of the sine function in Eq.(\ref{general stilde}), then there is an exact $\pi/2$ phase difference between the oscillations of $\tilde{s}^+(t)$ and $\tilde{j}^+(t)$, similar to the un-dampened LC circuit, in contrast to the over-damped case where $\tilde{j}^+$ follows adiabatically the dynamics of $\tilde{s}^+$.

\begin{figure}[h]
\includegraphics[width=85mm,scale=1]{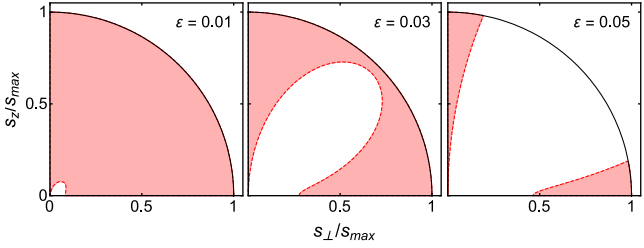}
\caption{\label{fig:AdiaN}(Colour online) The approximate demarcation of adiabatic from dabatic regions based on the condition $|\partial_t \vec{s}/\vec{s}| \sim \gamma$ (red shaded lines). The region of adiabaticity becomes smaller when the wave number of spin helix deviates away from $2q$.}
\end{figure}

The region of adiabaticity for various initial polarizations $s_{z,0}$ and $s_{\perp,0}$ are determined (approximately) by the condition $|\partial_t \vec{s}/\vec{s}| \sim \gamma$. It is shown in Fig.\ref{fig:AdiaN} that close to PSH, the adiabatic region prevails for most of the parameter regime except when the spin polarisation is small. On the other hand, as one moves away from PSH, the region of non-adiabatic evolution grows much larger. Starting from an arbitrary initial conditions, the spin dynamics might traverse both adiabatic and diabatic regimes and becomes much richer. In particular, close to the boundaries, it is necessary to deal with the full set of hydrodynamic equations (\ref{eq:DEOM},\ref{eq:SEOM},\ref{eq: current density EOM},\ref{eq: spin current density EOM}) that we derived before. 

{\em {Quenching of Rabi coupling on PSH.}} In the above analysis, we have assumed that the Rabi coupling is weak and can be neglected. Inclusion of Rabi term results in complex dynamics, for example an extra precession of spin density and spin current density in Eqs.(\ref{eq:SEOM},\ref{eq: spin current density EOM}). In the equilibrium state, due to the breaking of the emergent SU(2) symmetry, PSH is no longer stable and decays with a rate that is determined by $\Omega_R$. 

Considering the situation that the Rabi coupling is turned on suddenly at $t=0$ and remains fixed, all spin helices except the PSH with $k=2q$ vanish long before $t=0$. In the following the short-time effect of the Rabi coupling on the PSH will be studied. We can separate the densities and current densities into two parts, one from the PSH while the other from the leading correction due to Rabi coupling which vanishes for $t \leq 0$,
\begin{align}
n(x,t) =&~n_0 + \delta n(x,t)~,\\
\vec{s}(x,t) =&~\vec{s}_{\rm psh} + \delta \vec{s}(x,t)~,\\
j_0(x,t) =&~0 + \delta j_0(x,t)~,\\
\vec{j}(x,t) =&~0 + \delta \vec{j}(x,t)~.
\end{align}
Substituting the above expressions into the transport equations Eqs.(\ref{eq:DEOM},\ref{eq:SEOM},\ref{eq: current density EOM},\ref{eq: spin current density EOM}) and only keeping terms linear in small derivations and the Rabi coupling, we are led to an inhomogeneous diffusion equation of the form
\begin{equation}
\partial_t \vec{\delta V}(x,t) = \hat{\mathbf{H}}(x,\partial_x)\vec{\delta V}(x,t) + \vec{g}(x)~,
\end{equation}
with $\vec{\delta V} = (\delta n, \delta j_0, \delta s_z, \delta j_z, \delta s_x, \delta j_x, \delta s_y, \delta j_y)^T$ and $\vec{g}(x) = \Omega_R (0, {2q}s_{{\rm psh},y}/{m}, s_{{\rm psh},y},0,0,0, -s_{z,0},qn_0/(2m))^T$. The explicit form of $\hat{\mathbf{H}}(x,\partial_x)$ is given in Supplemental Material~\cite{supp} and the solution is given by
\begin{align} \label{deltaVSeries}
\vec{\delta V}(x,t)= \hat{\mathbf{H}}^{-1}\left[\exp({\hat{\mathbf{H}} t}) - \hat{\mathbf{I}}\right] \vec{g}(x).
\end{align}

To characterize the decay of transverse component of spin density due to Rabi coupling, we define a quantity that measures the amplitude of the spin helical structure 
\begin{equation}
R(t) = \dfrac{1}{|\vec{s}(x,0)|(\pi/q)}\int_{0}^{\frac{\pi}{q}}{\rm Re}[s^+(x,t) e^{-i2qx} ] \mathrm{d}x~,
\end{equation}
where $|\vec{s}(x,0)| = \sqrt{s_{\perp,0}^2 + s_{z,0}^2}$ is the initial spin magnitude. For $t>0$, Rabi coupling destroys the helical structure and results in decay of $R(t)$. The short-time behaviour is described by the leading terms of the series Eq.(\ref{deltaVSeries}). When $\Omega_R t \ll 1$,
\begin{equation} \label{LeadingR}
R(t) \approx \dfrac{s_{\perp,0}}{\sqrt{s_{\perp,0}^2 + s_{z,0}^2}}\left[ 1- \dfrac{(\Omega_R t)^2}{4} \right] + \mathcal{O}(t^4).
\end{equation}

%

{\em Experimental Considerations.} For $^{87}\mathrm{Rb}$ used in spin-orbit coupling experiment~\cite{nature09887}, the typical density is about ${n_0} =2\times10^{13}\,\mbox{cm}^{-3}$. For our calculation, we assume $T= 700\mbox{nK}$, well above the typical condensation temperature. The Raman laser defines a scale of wave number $k_L = (\sqrt{2}\pi/804.1\,\mbox{nm})/10$, chosen to be 10 times smaller than that in Ref.\cite{nature09887} to make the spin helical structure more visible. In the numerical calculations presented, we chose $q = 0.5 \hbar k_L$ and Rabi coupling $\hbar \Omega_R = 0.5 \hbar^2 k_L^2/(2m)$, appropriate to experimental situations. We assume the system is initially polarized with $|\vec{s}_0|=s_{\rm max}= {n_0}/2$. The intrinsic spin current relaxation rate $\gamma$ is chosen to be approximately $20$Hz, appropriate to $^{87}\mathrm{Rb}$~\cite{PhysRevLett.88.070403, PhysRevLett.88.230403}. Numerical simulations show that the time scale for spin dynamics in the adiabatic regime is of order of seconds, while it is of order of milliseconds in the diabatic regime, and can be observed experimentally. To initialize the system in a particular spin helical state, one can start with Rb atoms in the $\ket{\downarrow}$ state with no Raman lasers and apply a radio-frequency pulse to achieve a desired $s_z$ polarization. Afterwards, a small magnetic field with linear gradient $\Delta B$ can be applied to create the spin helical structure with wave vector $k$. To investigate the stability of spin helical structure, one can now turn on the Raman fields which create the spin-orbit coupling with strength $q$, and measure the evolution of transverse spin component.


\begin{acknowledgments}
This work is supported by Hong Kong Research Grants Council, GRF HKU 17305217, CRF C6026-16W, and the Croucher Foundation under the Croucher Innovation Award. 
\end{acknowledgments}


\bibliographystyle{apsrev4-1}

\end{document}



\title{Supplemental Material for \\
``Persistent Spin Helix and its Instability in Spin-orbit Coupled Normal Bose Gas''}


\author{Wai Ho Tang}
\affiliation{Department of Physics and Center of Theoretical and Computational Physics, The University of Hong Kong, Hong Kong, China}

\author{Shizhong Zhang}
\affiliation{Department of Physics and Center of Theoretical and Computational Physics, The University of Hong Kong, Hong Kong, China}

\date{\today}

\maketitle


\section{Derivation of Equations of Motion}
\subsection{Dynamics of Densities}
Given the Hamiltonian of system $\hat{\mathscr{H}}=\int d^3\mathbf{r}\sum_{\mu,\nu}\psi^\dagger_{\mu}(\mathbf{r})H_{\mu\nu}\psi_{\nu}(\mathbf{r})+\frac{1}{2}g\int d^3 \mathbf{r} \colon\hat{n}(\mathbf{r})\hat{n}(\mathbf{r})\colon$ with $\mu,\nu=\uparrow,\downarrow$ and
\begin{equation}
H_{\mu\nu}=\left[\dfrac{-\hbar^2 \nabla^2}{2m} + V(\mathbf{r})\right]\delta_{\mu\nu} -\frac{i\hbar q}{m}\sigma^z_{\mu\nu} \partial_x + \frac{\hbar\Omega_R}{2}\sigma^x_{\mu\nu}~.
\end{equation}
$\hat{\psi}_{\mu}(\mathbf{r})$ ($\hat{\psi}^\dagger_{\mu}(\mathbf{r})$) is the annihilation (creation) operator of a boson with spin component $\mu$ at position $\mathbf{r}$. $\sigma^i$ are the Pauli matrices. The dynamics of number density $\hat{n}(\mathbf{r})=\sum_{\mu}\hat{\psi}^\dagger_{\mu}(\mathbf{r})\hat{\psi}_{\mu}(\mathbf{r})$ and spin densities ${\hat{s}}_i(\mathbf{r}) = \frac{1}{2}\sum_{\mu,\nu}\hat{\psi}^\dagger_\mu(\mathbf{r})\sigma^i_{\mu\nu}\hat{\psi}_\nu(\mathbf{r})$ is governed by the equations of motion derived from Heisenberg's equation $i\hbar\braket{\partial_t \hat{A}} = \braket{[ \hat{A} , \hat{\mathscr{H}}]}$, explicitly
\begin{align}
\partial_t n + \nabla \cdot \mathbf{j}_0 &= -\dfrac{2}{m}\mathbf{q} \cdot \nabla s_z~,\\
\partial_t s_z + \nabla \cdot \mathbf{j}_z &= \Omega_R s_y - \dfrac{1}{2m}\mathbf{q} \cdot \nabla n~.\\
\partial_t s_x + \nabla \cdot \mathbf{j}_x &= - \dfrac{2}{\hbar}\mathbf{q} \cdot \mathbf{j}_y~,\\
\partial_t s_y + \nabla \cdot \mathbf{j}_y &= - \Omega_R s_z + \dfrac{2}{\hbar}\mathbf{q} \cdot \mathbf{j}_x~,
\end{align}
with spin-orbit coupling $\mathbf{q} = q \hat{\mathbf{x}}$ and the help of Wick's theorem:
\begin{equation}
\braket{\hat{\psi}^\dagger_\alpha\,\hat{\psi}^\dagger_\beta\,\hat{\psi}_m\,\hat{\psi}_n} = \braket{\hat{\psi}^\dagger_\alpha\,\hat{\psi}_m}\braket{\hat{\psi}^\dagger_\beta\,\hat{\psi}_n} + \braket{\hat{\psi}^\dagger_\alpha\,\hat{\psi}_n}\braket{\hat{\psi}^\dagger_\beta\,\hat{\psi}_m} + \braket{\hat{\psi}^\dagger_\alpha\,\hat{\psi}^\dagger_\beta}\braket{\hat{\psi}_m\,\hat{\psi}_n}~,
\end{equation}
the last term called anomalous term is omitted in the present case of normal phase. The number current $\mathbf{j}_0=\langle\hat{\mathbf{j}}_0\rangle$ with $\hat{\mathbf{j}}_0=-{i\hbar}/({2m})\sum_{\mu}(\hat{\psi}^\dagger_{\mu}\nabla \hat{\psi}_{\mu}-\nabla\hat{\psi}^\dagger_{\mu}\,\hat{\psi}_{\mu})$ and spin currents $\mathbf{j}_i =\langle\hat{\mathbf{j}}_i\rangle$ with $\hat{\mathbf{j}}_i=-{i\hbar}/({4m})\sum_{\mu,\nu}\hat{\psi}^\dagger_\mu \sigma^i_{\mu\nu}\,\nabla \hat{\psi}_\nu + \mathrm{H.c.}$ are defined in \textit{conventional} manner, i.e. only taking kinetic term of Hamiltonian into account. The equations of motion of these conventional currents are necessary to complete the system's dynamics.

\subsection{Dynamics of Current Densities}
After straightforward but tedious calculation, equations of motion of conventional number and spin current densities are obtained,
\begin{align}
\partial_t j_{0,\alpha} + \partial_\beta \braket{\hat{\Pi}^0_{\alpha\beta}} &= -\dfrac{n}{m} \partial_\alpha V(\mathbf{r}) -\dfrac{3g}{4m}\partial_\alpha(n^2) - \dfrac{g}{m}\partial_\alpha(s_x^2 + s_y^2 + s_z^2 \,) -\dfrac{2(\mathbf{q}\cdot\nabla)}{m} j_{z,\alpha}~,\\
\partial_t j_{z,\alpha} + \partial_\beta \braket{\hat{\Pi}^z_{\alpha\beta}} &= -\dfrac{s_z}{m}\partial_\alpha V(\mathbf{r}) + \dfrac{2g}{\hbar}(s_x j_{y,\alpha} - s_y j_{x,\alpha}) - \dfrac{3g}{2m} s_z \partial_\alpha n - \dfrac{g}{2m} n \partial_\alpha s_z -\dfrac{(\mathbf{q}\cdot\nabla)}{2m} j_{0,\alpha} + \Omega_R j_{y,\alpha}~,\\
\partial_t j_{x,\alpha} + \partial_\beta \braket{\hat{\Pi}^x_{\alpha\beta}} &= -\dfrac{s_x}{m}\partial_\alpha V(\mathbf{r}) + \dfrac{2g}{\hbar}(s_y j_{z,\alpha} - s_z j_{y,\alpha}) - \dfrac{3g}{2m} s_x \partial_\alpha n - \dfrac{g}{2m} n \partial_\alpha s_x -\dfrac{2q_\beta \braket{\hat{\Pi}^y_{\alpha\beta}}}{\hbar}~,\\
\partial_t j_{y,\alpha} + \partial_\beta \braket{\hat{\Pi}^y_{\alpha\beta}} &= -\dfrac{s_y}{m}\partial_\alpha V(\mathbf{r}) + \dfrac{2g}{\hbar}(s_z j_{x,\alpha} - s_x j_{z,\alpha}) - \dfrac{3g}{2m} s_y \partial_\alpha n - \dfrac{g}{2m} n \partial_\alpha s_y + \dfrac{2q_\beta \braket{\hat{\Pi}^x_{\alpha\beta}}}{\hbar} -\Omega_R j_{z,\alpha}~.
\end{align}
Repeated index $\beta$ means summation of $\beta = x,y,z$. The generalized momentum flux tensors $\hat{\Pi}^i_{\alpha\beta}$ are defined as
\begin{align}
\hat{\Pi}^0_{\alpha\beta} &= \left(\dfrac{\hbar}{2m}\right)^2 (\partial_\alpha \hat{\psi}_\uparrow^\dagger\,\partial_\beta\hat{\psi}_\uparrow - \hat{\psi}_\uparrow^\dagger\,\partial_\alpha\partial_\beta\hat{\psi}_\uparrow + \partial_\alpha \hat{\psi}_\downarrow^\dagger\,\partial_\beta\hat{\psi}_\downarrow - \hat{\psi}_\downarrow^\dagger\,\partial_\alpha\partial_\beta\hat{\psi}_\downarrow) + \mathrm{H.c.}~,\\
\hat{\Pi}^z_{\alpha\beta} &= \dfrac{1}{2}\left(\dfrac{\hbar}{2m}\right)^2 (\partial_\alpha \hat{\psi}_\uparrow^\dagger\,\partial_\beta\hat{\psi}_\uparrow - \hat{\psi}_\uparrow^\dagger\,\partial_\alpha\partial_\beta\hat{\psi}_\uparrow - \partial_\alpha \hat{\psi}_\downarrow^\dagger\,\partial_\beta\hat{\psi}_\downarrow + \hat{\psi}_\downarrow^\dagger\,\partial_\alpha\partial_\beta\hat{\psi}_\downarrow) + \mathrm{H.c.}~,\\
\hat{\Pi}^x_{\alpha\beta} &= \dfrac{1}{2}\left(\dfrac{\hbar}{2m}\right)^2 (\partial_\alpha \hat{\psi}_\uparrow^\dagger\,\partial_\beta\hat{\psi}_\downarrow - \hat{\psi}_\uparrow^\dagger\,\partial_\alpha\partial_\beta\hat{\psi}_\downarrow + \partial_\beta \hat{\psi}_\uparrow^\dagger\,\partial_\alpha\hat{\psi}_\downarrow - \partial_\alpha\partial_\beta\hat{\psi}_\uparrow^\dagger\,\hat{\psi}_\downarrow) + \mathrm{H.c.}~,\\
\hat{\Pi}^y_{\alpha\beta} &= \dfrac{1}{2i}\left(\dfrac{\hbar}{2m}\right)^2 (\partial_\alpha \hat{\psi}_\uparrow^\dagger\,\partial_\beta\hat{\psi}_\downarrow - \hat{\psi}_\uparrow^\dagger\,\partial_\alpha\partial_\beta\hat{\psi}_\downarrow + \partial_\beta \hat{\psi}_\uparrow^\dagger\,\partial_\alpha\hat{\psi}_\downarrow - \partial_\alpha\partial_\beta\hat{\psi}_\uparrow^\dagger\,\hat{\psi}_\downarrow) + \mathrm{H.c.}~.
\end{align}
In contrast to Oktel's work where the densities is nearly uniform over the time duration studied, spin helix having large wave number $k \sim 2q/\hbar$ implies that the gradient terms associated with exchange interaction are comparable with the Leggett-Rice terms, hence these gradient terms should be retained in present studies.

\subsection{Quasi-one-dimensional limit}
The calculation of ensemble average of momentum flux tensors are not trivial in presence of spin-orbit coupling. Only the component $(\alpha\beta)=(xx)$ of the tensors are concerned as quasi-one-dimensional limit is taken below. In the case without Rabi coupling, the number and longitudinal spin densities are conserved so their corresponding current densities are well defined. These currents are modified by spin-orbit coupling as shown in main text. Their modifications can also be obtained by performing gauge transformation, which eliminates the spin-orbit coupling at the expense of introducing space-dependent spin quantisation axis,
\begin{equation}
\tilde{\psi}=\exp(i q x \sigma_z)\psi~.
\end{equation}
 Therefore the gradient expansion of momentum flux tensors $\braket{\Pi^0_{xx}}$ and $\braket{\Pi^z_{xx}}$ could be obtained by first carrying out the standard gradient expansion on energy momentum flux without SOC ($\tilde{\Pi}^0_{xx}$, $\tilde{\Pi}^z_{xx}$) and then using the gauge transformation to obtain that for $\braket{\Pi^0_{xx}}$ and $\braket{\Pi^z_{xx}}$. As a result, we obtain,
\begin{align}
\braket{\Pi^0_{xx}} &= \left(\dfrac{k_B T}{m} - \dfrac{q^2}{m^2} \right)n - \dfrac{4q}{m} j_z~,\\
\braket{\Pi^z_{xx}} &= \left(\dfrac{k_B T}{m} - \dfrac{q^2}{m^2} \right)s_z - \dfrac{q}{m} j_0~.
\end{align}
However, the transverse spin components in general does not satisfy the equations of continuity owing to spin-orbit coupling, as a result, the definition of the corresponding energy-momentum tensors is not uniquely fixed. In our study, we recognise that the ``streaming'' term (group velocity) is not modified in the $\hat{x}$ and $\hat{y}$-directions and as a result, we shall apply the standard gradient expansion to their corresponding energy momentum flux tensor,
\begin{align}
\braket{\Pi^x_{xx}} &= \dfrac{k_B T}{m} s_x~,\\
\braket{\Pi^y_{xx}} &= \dfrac{k_B T}{m} s_y~.
\end{align}

Since the external potential $V(\mathbf{r})$ is uniform along $\hat{\mathbf{x}}$-direction but has strong confinement along $\hat{\mathbf{y}}$-$\hat{\mathbf{z}}$ plane, the system is in quasi-one-dimensional limit, in which $n(x,y,z)$ is replaced by the central peak value $n(x,0,0)$ (similarly for other quantities), the Gaussian distribution profile along $\hat{\mathbf{y}}$-$\hat{\mathbf{z}}$ plane is averaged out and results in replacement of interaction strength $g \to g/2$. After adopting redefinitions of currents defined in main text, one have the equations of motion in quasi-one-dimensional limit,
\begin{align}
\partial_t n + \partial_x j_0 &= 0~,\label{eq n} \\
\partial_t \vec{s} + \partial_x \vec{j} &= \Omega_R \hat{x}\times\vec{s} + (2q/\hbar)\hat{z}\times\vec{j}~,\label{eq s} \\
\partial_t j_0 + \dfrac{k_B T}{m} \partial_x n &=\frac{2q}{m}\Omega_R s_y -\frac{g}{2m}\partial_x\left(\frac{3}{4}n^2+\vec{s}^{\,2}\right)~,\label{eq j0} \\
\partial_t \vec{j} + \alpha \partial_x \vec{s} &= \left(\Omega_R \hat{x} + \frac{g}{\hbar} \vec{s}\right)\times\vec{j} +\frac{2q\alpha}{\hbar}\hat{z}\times\vec{s}+\frac{qn\Omega_R}{2m}\hat{y} - \frac{3g}{4m} (\partial_x n)\vec{s} - \gamma \vec{j}~,\label{eq spin j}
\end{align}
where $\alpha = k_B T/m + ng/(4m)$, the relaxation of spin current $-\gamma \vec{j}$ is phenomenologically added. Vectorial quantities are in spin space.

\section{Adiabatic Evolution of Spin helix}
Hereafter $\hbar = 1$. The spin dynamics of spin helix simplifies and given by Eqs.(7-9) in main text, explicitly
\begin{align}
\partial_t \tilde{s}^+ &= - i (k-2q)\tilde{j}^+~,\label{EOM1}\\
\partial_t \tilde{j}^+  &= (i \lambda s_{z,0} - \gamma) \tilde{j}^+ -i [\alpha(k-2q) + \lambda j_z]\tilde{s}^+,\label{EOM2}\\
\partial_t j_z &= \lambda {\rm Im}[\tilde{s}^- \tilde{j}^+] -\gamma j_z.\label{EOM3}
\end{align}
For $k \neq 2q$, there is no persistent spin helix and the spin dynamics could be classified into two regimes. In this section the adiabatic regime where the spin density varies slowly is discussed. The diabatic regime where the spin density is changing fast will be the subject of the next section.

The dynamical equation of spin current is inhomogeneous so the spin current consists of a homogeneous solution and an inhomogeneous solution. The former dies out in time scale about $\gamma^{-1}$ and is unimportant, while the latter is what we desire. The spin current will acquire the inhomogeneous value in the duration $\Delta t \sim \gamma^{-1}$, which is obtained by setting $\partial_t \vec{j} = 0$. The solution is valid only when every terms in the equation are slowly varying in the time interval $\Delta t \sim \gamma^{-1}$, i.e. $|\partial_t \vec{s}/\vec{s}\,| \lesssim \gamma$, the validity will be justified at the end.

Adiabatic assumption $\partial_t \tilde{j}^+ = \partial_t j_z = 0$ implies
\begin{align}
j_z &= \dfrac{\lambda}{\gamma} {\rm Im}[\tilde{s}^- \tilde{j}^+]~,\label{adia jz}\\
\tilde{j}^+ &= \dfrac{-i\tilde{s}^+}{\gamma - i \lambda s_{z,0}}\left[\alpha \left(k-2q\right) + \lambda j_z \right]~.\label{adia j+}
\end{align}
By writing $\tilde{s}^{\pm} (t) \equiv s_\perp(t) e^{\pm i\theta(t)}$, $j_z$ could be solved by substituting Eq.(\ref{adia j+}) into Eq.(\ref{adia jz}), and hence $\tilde{j}^+$:
\begin{align}
j_z &= -\dfrac{\lambda s_\perp^2 \alpha(k-2q)}{\gamma^2 + \lambda^2 (s_\perp^2 + s_{z,0}^2)}~,\\
\tilde{j}^+ &= \tilde{s}^+ \dfrac{\alpha(k-2q)(\lambda s_{z,0} - i \gamma)}{\gamma^2 + \lambda^2 (s_\perp^2 + s_{z,0}^2)}~.
\end{align}
As $\partial_t \tilde{s}^+ = e^{i\theta}[\partial_t s_\perp + i(s_\perp \partial_t \theta)]$, separation of real and imaginary parts results in the rates of change of the transverse spin magnitude $s_\perp (t)$ and phase $\theta(t)$,
\begin{align}
\partial_t s_\perp &= -s_\perp \dfrac{\gamma \alpha \left(k-2q\right)^2}{\gamma^2 + \lambda^2 (s_\perp^2 + s_{z,0}^2)}~,\\
\partial_t \theta &= \dfrac{\lambda s_{z,0}}{\gamma}\dfrac{\partial_t s_\perp}{s_\perp}~.
\end{align}
Direct integration implies algebraic equations of $s_\perp (t)$ and $\theta(t)$,
\begin{align}
(\gamma^2 + \lambda^2 s_{z,0}^2)\ln\dfrac{s_\perp(t)}{s_{\perp,0}} + \frac{\lambda^2}{2}\left[s_{\perp}^2(t) - s_{\perp,0}^2\right] = -\alpha (k - 2q)^2 \gamma t~,\\
\theta(t) = \dfrac{\lambda s_{z,0}}{\gamma}\ln \dfrac{s_\perp(t)}{s_{\perp,0}}~.
\end{align}

Above solution is valid only when it is consistent with the adiabatic assumption,
\begin{align}
|\partial_t \vec{s}/\vec{s}\,|\, = \sqrt{\dfrac{(\partial_t s_\perp)^2 + (s_\perp \partial_t \theta)^2}{s_\perp^2 + s_{z,0}^2}} \lesssim \gamma~.
\end{align}
For PSH $k = 2q$ this condition is always satisfied as the left hand side vanishes. For $k = (1+\varepsilon)2q \neq 2q$ the condition becomes
\begin{equation}
\left( k- 2q \right)^2 = \varepsilon^2 (2q)^2 \lesssim \dfrac{\gamma}{\sqrt{\gamma^2 + \lambda^2 s_{z,0}^2}} \dfrac{\sqrt{s_\perp^2 + s_{z,0}^2}}{s_\perp} \dfrac{\gamma^2 + \lambda^2(s_\perp^2 + s_{z,0}^2)}{\alpha}~.
\end{equation}
In conclusion, adiabatic regimes is valid only in the vicinity of PSH, i.e. small $\varepsilon$.

\section{Diabatic Evolution of Spin helix}
We have seen that the adiabatic condition is only valid in the vicinity of PSH (say $|\varepsilon| \lesssim 0.05$). This precision is hard to control in experiments, it means we have to study the general regime as well. As previous, we only have interest in the dynamics of spin helix, which is governed by Eqs.(\ref{EOM1}-\ref{EOM3}). Attention should be paid to Eq.(\ref{EOM2}), when $k$ is far away from $2q$, the time-dependent term $j_z(t)$ is dominated by the term with $\alpha$, this neglect justified below benefits us that the coefficients are time-independent and that the dynamics of $\tilde{s}^+$ and $\tilde{j}^+$ are closed. Once the transverse spin and current are known, the longitudinal current $j_z(t)$ can be solved by direct integration on Eq.(\ref{EOM3}).

The equations of motion of spin helix can be rewritten as below,
\begin{align}
\partial_t \left[ \begin{array}{cc}
\tilde{s}^+ \\ \tilde{j}^+ \end{array} \right]
&= \left[ \begin{array}{cc}
0 & -i(k-2q) \\
-i[\alpha(k-2q) + \lambda j_z] & i \lambda s_{z,0} - \gamma
\end{array} \right]
\left[ \begin{array}{cc}
\tilde{s}^+ \\ \tilde{j}^+ \end{array} \right]
= \Lambda \left[ \begin{array}{cc}
\tilde{s}^+ \\ \tilde{j}^+ \end{array} \right]~.\label{eigen1}
\end{align}
The time dependence of $j_z$ results in that of eigenvalues and eigenstates. We approximate $j_z$ by its equilibrium value, $j_z(t\to \infty) = 0$ thus the eigenvalues and eigenvectors become time independent, the general solution can be decomposed into the form
\begin{equation}
\left[ \begin{array}{cc}
\tilde{s}^+(t) \\ \tilde{j}^+(t) \end{array} \right]
=
C_1\,e^{\Lambda_1 t} \left[ \begin{array}{cc}
\tilde{s}^+ \\ \tilde{j}^+ \end{array} \right]_1
+
C_2\,e^{\Lambda_2 t} \left[ \begin{array}{cc}
\tilde{s}^+ \\ \tilde{j}^+ \end{array} \right]_2~,
\end{equation}
where constants $C_{1,2}$ are fixed by the initial condition $(\tilde{s}^+, \tilde{j}^+)_{t=0} = (s_{\perp,0},0)$. Vanishing initial current is assumed for convenience but it has captured enough features. Therefore the transverse components of spin density and spin current are known,
\begin{align}
s^+(x,t) &= \dfrac{s_{\perp,0}}{2(\Gamma_1 + i \Gamma_2)} \exp\left( -\dfrac{\gamma t}{2} + i \dfrac{\lambda s_{z,0}}{2} t + ikx \right) \times \left\{ (\gamma - i \lambda s_{z,0})\sin[(\Gamma_1 + i \Gamma_2)t] + 2(\Gamma_1 + i \Gamma_2)\cos[(\Gamma_1 + i \Gamma_2)t]\right\}~,\\
j^+(x,t) &= \dfrac{s_{\perp,0}}{2(\Gamma_1 + i \Gamma_2)} \exp\left( -\dfrac{\gamma t}{2} + i \dfrac{\lambda s_{z,0}}{2} t + ikx \right) \times \left\{-i2\alpha(k-2q) \sin[(\Gamma_1 + i \Gamma_2)t]\right\}~,
\end{align}
where
\begin{equation}
\Gamma_1 + i \Gamma_2 = \sqrt{\alpha \left(k-2q\right)^2 - \left(\dfrac{\gamma}{2} - i\dfrac{\lambda s_{z,0}}{2}\right)^2}~.
\end{equation}
The longitudinal component of spin current obeying Eq.(\ref{EOM3}) with initial condition $j_z(0) = j_{z,0}$ is given by
\begin{equation}
j_z(t) = j_{z,0} e^{-\gamma t} - e^{-\gamma t} \dfrac{\lambda s_{\perp,0}^2 \alpha (k-2q)}{4(\Gamma_1^2 + \Gamma_2^2)} \left[ \cosh(2\Gamma_2 t) + \dfrac{\gamma}{2}\dfrac{\sinh(2\Gamma_2 t)}{\Gamma_2} - \cos(2\Gamma_1 t) - \dfrac{\gamma}{2}\dfrac{\sin(2\Gamma_1 t)}{\Gamma_1} \right]~.
\end{equation}

If the time dependence of $j_z(t)$ is retained, the eigenvalues of Eq.(\ref{eigen1}) become
\begin{equation}
\Lambda_{1,2} = -\dfrac{\gamma}{2} + i \dfrac{\lambda s_{z,0}}{2} \pm \sqrt{\alpha \left(k-2q\right)^2 - \left(\dfrac{\gamma}{2} - i\dfrac{\lambda s_{z,0}}{2}\right)^2 + \lambda j_z \left(k - 2q\right)}~.
\end{equation}
The first two terms remain unchanged but the square root term $\Gamma_1 + i \Gamma_2$ becomes time dependent. Estimate of magnitude shows that
\begin{equation}
\alpha \left(k-2q\right)^2 ~\gg~ \lambda |j_z| |k - 2q| ~\sim~ |\dfrac{\gamma}{2} - i\dfrac{\lambda s_{z,0}}{2}|^2~,\label{magnitude}
\end{equation}
hence the approximation $\Gamma_1 + i \Gamma_2 \approx \sqrt{\alpha} |k-2q| = \Gamma$ simplifies the expressions,
\begin{align}
s^+(x,t) &= s_{\perp,0} \exp\left(-\dfrac{\gamma t}{2} + i \dfrac{\lambda s_{z,0}}{2} t + ikx \right) \left( \cos \Gamma t + \dfrac{\gamma - i \lambda s_{z,0}}{2\Gamma}\sin \Gamma t \right)~,\\
j^+(x,t) &= i \sqrt{\alpha} s_{\perp,0} \exp\left(-\dfrac{\gamma t}{2} + i \dfrac{\lambda s_{z,0}}{2} t + ikx \right) \sin(\Gamma t)\,\mbox{sign}(2q - k)~,\\
j_z(t) &= j_{z,0} e^{-\gamma t} - e^{-\gamma t} \dfrac{\lambda s_{\perp,0}^2}{4 (k-2q)} \left[ 1 + \gamma t - \cos(2\Gamma t) - \dfrac{\gamma}{2}\dfrac{\sin(2\Gamma t)}{\Gamma} \right]~.
\end{align}

Recalling that $\lambda |j_z| \ll \Gamma$ is assumed when de-coupling the equations of motion, for self-consistency we substitute the amplitude of $j_z$
\begin{equation}
\lambda |j_z|/\Gamma \sim (\lambda s_{\perp,0}/\Gamma)^2 \ll 1~,
\end{equation}
the last inequality is identical to the limit taken in Eq.(\ref{magnitude}). 

\section{Quenching of Rabi coupling on PSH}
After substituting Eqs.(20-23) in main text into Eqs.(\ref{eq n}-\ref{eq spin j}) and dropping quadratic terms of fluctuations, one can obtain an inhomogeneous diffusion equation for the dynamics of fluctuation $\vec{\delta V} = (\delta n, \delta j_0, \delta s_z, \delta j_z, \delta s_x, \delta j_x, \delta s_y, \delta j_y)^T$,
\begin{equation}
\partial_t \vec{\delta V}(x,t) = \hat{\mathbf{H}}(x,\partial_x)\vec{\delta V}(x,t) + \vec{g}(x)
\end{equation}
with source term $\vec{g}(x) = \Omega_R (0, {2q}s_{{\rm psh},y}/{m}, s_{{\rm psh},y},0,0,0, -s_{z,0},qn_0/(2m))^T$. The evolution matrices are given by
\begin{align}
\hat{\mathbf{H}} &= \left[
\begin{array}{cc}
H_{11} & H_{12} \\
H_{21} & H_{22} \\
\end{array}
\right]~,\\
H_{11} &= \left[
\begin{array}{cccc}
0 & -\partial_x & 0 & 0 \\
-\left(\frac{k_B T}{m} + \frac{3g n_0}{4m}\right) \partial_x & 0 & -\frac{g}{m}(\partial_x s_{z,0} + s_{z,0} \partial_x) & 0 \\
0 & 0 & 0 & -\partial_x \\
-\frac{3g s_{z,0}}{4m}\partial_x & 0 & -\alpha \partial_x & -\gamma
\end{array}
\right]~,\\
H_{12} &= \left[
\begin{array}{cccc}
0 & 0 & 0 & 0 \\
-\frac{g}{m}(\partial_x s_{{\rm psh},x} + s_{{\rm psh},x} \partial_x) & 0 & -\frac{g}{m}(\partial_x s_{{\rm psh},y} + s_{{\rm psh},y} \partial_x) + \frac{2q\Omega_R}{m} & 0 \\
0 & 0 & +\Omega_R & 0 \\
0 & -\lambda s_{{\rm psh},y} & 0 & + \lambda s_{{\rm psh},x} + \Omega_R
\end{array}
\right]~,\\
H_{21} &= \left[
\begin{array}{cccc}
0 & 0 & 0 & 0 \\
-\frac{3g s_{{\rm psh},x}}{4m}\partial_x & 0 & 0 & +\lambda s_{{\rm psh},y} \\
0 & 0 & -\Omega_R & 0 \\
-\frac{3g s_{{\rm psh},y}}{4m}\partial_x + \frac{q\Omega_R}{2m} & 0 & 0 & -\lambda s_{{\rm psh},x} - \Omega_R \\
\end{array}
\right]~,\\
H_{22} &= \left[
\begin{array}{cccc}
0 & -\partial_x & 0 & -2q \\
-\alpha\partial_x & -\gamma & - 2q\alpha & -\lambda s_{z,0} \\
0 & 2q & 0 & -\partial_x \\
2q\alpha & \lambda s_{z,0} & -\alpha \partial_x & -\gamma
\end{array}
\right]~.
\end{align}
The inhomogeneous diffusion equation can be solved by repeating integrations,
\begin{align}
\vec{\delta V}(x,t) &= \int_{0}^{t} \vec{g}(x)\,\mathrm{d}t^\prime + \int_{0}^{t} \hat{\mathbf{H}}(x,\partial_x) \vec{\delta V}(x,t^\prime)\,\mathrm{d}t^\prime \nonumber \\
&= \int_{0}^{t} \vec{g}(x)\,\mathrm{d}t^\prime + \int_{0}^{t} \int_{0}^{t^\prime} \hat{\mathbf{H}}(x,\partial_x) \vec{g}(x)\,\mathrm{d}t^{\prime \prime} \mathrm{d}t^\prime + \int_{0}^{t} \int_{0}^{t^\prime} \hat{\mathbf{H}}^2 \vec{\delta V}(x, t^{\prime\prime})\,\mathrm{d}t^{\prime \prime} \mathrm{d}t^\prime \nonumber \\
&= \cdots = \left(\sum_{n=1}^{\infty} \dfrac{t^n}{n!} \hat{\mathbf{H}}^{n-1} \right)\vec{g}(x) = \hat{\mathbf{H}}^{-1}\left[\exp(\hat{\mathbf{H}} t) - \hat{\mathbf{I}} \right] \vec{g}(x)~,
\end{align}
and imposing initial condition $\vec{\delta V}(x,0) = 0$, i.e. only the inhomogeneous solution is concerned. Identical solution could be obtained alternatively by letting the form of $\vec{\delta V}(x,t)$ as
\begin{equation}
\vec{\delta V}(x,t) = \hat{\mathbf{S}}(t)\vec{g}(x)~~\mbox{with}~~\hat{\mathbf{S}}(0)=0~,
\end{equation}
therefore the diffusion equation becomes
\begin{align}
\partial_t [\hat{\mathbf{S}}(t)\vec{g}(x)] &= \hat{\mathbf{H}}(x,\partial_x) [\hat{\mathbf{S}}(t)\vec{g}(x)] + \vec{g}(x)\\
\partial_t \hat{\mathbf{S}}(t) &= \hat{\mathbf{H}}(x,\partial_x) \hat{\mathbf{S}}(t) + \hat{\mathbf{I}}~,
\end{align}
and implies
\begin{equation}
\hat{\mathbf{S}}(t) = \sum_{n=1}^{\infty} \dfrac{t^n}{n!} \hat{\mathbf{H}}^{n-1} = \hat{\mathbf{H}}^{-1}\left[\exp(\hat{\mathbf{H}} t) - \hat{\mathbf{I}} \right]~.
\end{equation}
Above solution is valid only for $\vec{\delta V} \approx \vec{g}(x)t \ll \vec{V}_0$, i.e. $\Omega_R t \ll 1$.